\documentclass[twocolumn,aps,prl,superscriptaddress]{revtex4}
\usepackage{amsmath}
\usepackage{amssymb}
\usepackage{color}
\usepackage[dvips]{graphicx}

\def\textbf#1{{\bf #1}}
\def\be{\begin{equation}}
\def\ee{\end{equation}}
\def\ben{\begin{eqnarray}}
\def\een{\end{eqnarray}}
\def\eea{\end{array}}
\def\bea{\begin{array}}

\newcommand{\bei}{\begin{itemize}}
\newcommand{\eei}{\end{itemize}}
\newcommand{\ket}[1]{|#1\rangle}

\usepackage{color}

\begin{document}

\title{Experimental estimation of the dimension of classical and quantum systems}

\author{Martin Hendrych}
\affiliation{ICFO-Institut de Ci\`{e}ncies Fot\`{o}niques, 08860
Castelldefels, Barcelona, Spain}
\email{martin.hendrych@icfo.es}
\author{Rodrigo Gallego}
\affiliation{ICFO-Institut de Ci\`{e}ncies Fot\`{o}niques, 08860
Castelldefels, Barcelona, Spain}
\author{Michal Mi\v{c}uda}
\affiliation{Department of Optics, Palack\'y University,
          17.~listopadu 12, 77146 Olomouc, Czech Republic}
\affiliation{ICFO-Institut de Ci\`{e}ncies Fot\`{o}niques, 08860
Castelldefels, Barcelona, Spain}
\author{Nicolas Brunner}
\affiliation{H.H. Wills Physics Laboratory, University of Bristol, Bristol, BS8 1TL, United Kingdom}
\author{Antonio Ac\'in}
\affiliation{ICFO-Institut de Ci\`{e}ncies Fot\`{o}niques, 08860
Castelldefels, Barcelona, Spain} \affiliation{ICREA-Instituci\'o
Catalana de Recerca i Estudis Avan\c cats, Llu\'{i}s Companys 23,
08010 Barcelona, Spain}
\author{Juan P. Torres}
\affiliation{ICFO-Institut de Ci\`{e}ncies Fot\`{o}niques, 08860
Castelldefels, Barcelona, Spain}
\affiliation{Dept. Signal Theory
and Communications, Universitat Polit\`{e}cnica de Catalunya,
08034 Barcelona, Spain}



\begin{abstract}
An overwhelming majority of experiments in classical and quantum
physics make a priori assumptions about the dimension of the
system under consideration. However, would it be possible to
assess the dimension of a completely unknown system only from the
results of measurements performed on it, without any extra
assumption? The concept of a dimension
witness~\cite{brunner,perezgarcia,wehner,perez,gbha} answers this
question, as it allows one to bound the dimension of an unknown
classical or quantum system in a device-independent manner, that
is, only from the statistics of measurements performed on it.
Here, we report on the experimental demonstration of dimension
witnesses in a prepare and measure scenario~\cite{gbha}. We use
pairs of photons entangled in both polarization and orbital
angular momentum \cite{molina-terriza2007,mair2001} to generate
ensembles of classical and quantum states of dimensions up to 4.
We then use a dimension witness to certify their dimensionality as
well as their quantum nature. Our results open new avenues for the
device-independent estimation of unknown quantum
systems~\cite{mayers,bardyn,bancal,rabelo} and for applications in
quantum information science~\cite{marcin,guo}.
\end{abstract}

\maketitle

Any physical theory aims at providing an explanation for the
results of measurements performed on a system under different
conditions. Given a system, one makes some general and plausible
assumptions about its behavior, builds a theory compatible with
these assumptions and tests its predictive power under different
experimental configurations. The dimension of the system, that is,
the number of relevant and independent degrees of freedom needed
to describe it, is in general an a priori parameter and represents
one of these initial assumptions. In general, the failure of a
theoretical model in predicting experimental data does not
necessarily imply that the assumption on the dimensionality is
incorrect, as there might exist a different model in a space of
the same dimension able to reproduce the observed data.

A natural question is whether this approach can be reversed and
whether the dimension of an unknown system, classical or quantum,
can be estimated experimentally. Clearly, the best one can hope
for is to provide lower bounds on this unknown dimension. Indeed,
one can never exclude that more degrees of freedom are necessary
to describe the system in more complex experimental arrangements.
The goal, then, is to obtain a lower bound on the dimension of the
unknown system from the observed measurement data without making
any assumption about the detailed functioning of the devices used
in the experiment. Besides its fundamental interest, estimating
the dimension of an unknown quantum system is also relevant from
the perspective of quantum information, where the Hilbert space
dimension is considered as a resource (see, e.g., \cite{lanyon}).
For instance, the optimal quantum realization of some protocols,
e.g. bit commitment, requires systems of a minimal dimension
\cite{BC}. Moreover, the dimension of quantum systems plays a
crucial role in security proofs of standard quantum key
distribution schemes that become insecure if the dimension is
higher than assumed~\cite{acingisin}.

The concept of a \textit{dimension witness} allows one to
establish lower bounds on the dimension of an unknown system in a
device-independent way, that is, only from the collected
measurement statistics. It was first introduced for quantum
systems in connection with Bell inequalities in~\cite{brunner},
and further developed
in~\cite{perezgarcia,vertesi1,vertesi2,vertesi4,junge,briet,junge2}.
Other techniques to estimate the dimension have been developed in
scenarios involving random access codes~\cite{wehner}, or the time
evolution of a quantum observable~\cite{perez}.

\begin{figure}[t]
  \includegraphics[width=0.8\columnwidth]{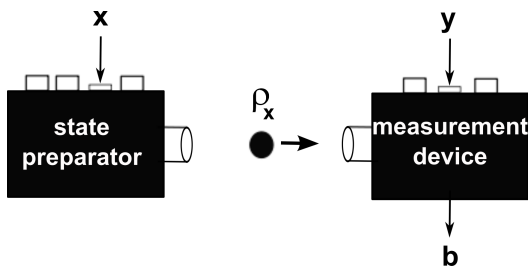}
  \caption{
 Sketch of a device-independent test of the dimension of an ensemble of quantum states.
 Our scenario features two black boxes: a state preparator, which generates one of $n$ possible states on demand,
 and a measurement device which performs one of $m$ possible measurements on the state generated.}
 \label{figure1}
\end{figure}

More recently, a general framework for the study of this question
has been proposed in~\cite{gbha}. In this approach, dimension
witnesses are defined in a prepare and measure scenario where an
unknown system is subject to different preparations and
measurements. One of the advantages of this approach is its
simplicity from an experimental viewpoint when compared to
previous proposals. The considered scenario consists of two
devices (see Fig.~\ref{figure1}), the {\it state preparator} and
the {\it measurement device}. These devices are seen as black
boxes, as no assumptions are made on their functioning. The state
preparator prepares upon request a state. The box features $N$
buttons which label the prepared state; when pressing button $x$,
the box emits a state $\rho_x$, where $x \in \{1,...,n\}$. The
prepared state is then sent to the measurement device. The
measurement box performs a measurement $y \in \{1,...,m\}$ on the
state, delivering outcome $b \in \{1,...,k\}$. The experiment is
thus described by the probability distribution $P(b|x,y)$, giving
the probability of obtaining outcome $b$ when measurement $y$ is
performed on the prepared state $\rho_x$. The goal is to estimate
the minimal dimension that the ensemble $\left\{ \rho_x \right\}$
must have to be able to describe the observed statistics.
Moreover, for a fixed dimension, we also aim at distinguishing
sets of probabilities $P(b|x, y)$ that can be obtained from
ensembles of quantum states, but not from ensembles of classical
states. This allows one to guarantee the quantum nature of an
ensemble of states under the assumption that the dimensionality is
bounded.

Formally, a probability distribution $P(b|x,y)$ admits a
$d$-dimensional representation if it can be written in the form
\begin{equation}\label{Q}
P(b|x,y)=\textrm{tr}(\rho_{x} M_{b}^{y}),
\end{equation}
for some state $\rho_x$ and operator $M_b^y$ acting on
$\mathbb{C}^d$. We then say that $P(b|x,y)$ has a classical
$d$-dimensional representation if any state of the ensemble
$\left\{ \rho_x \right\}$ is a classical state of dimension $d$,
i.e., a probability distribution over classical dits (or
equivalently, all the states of the ensemble commute).

A dimension witness for classical (quantum) systems of dimension
$d$ is defined by a linear combination of the observed
probabilities $P(b|x,y)$, defined by a tensor of real coefficients
$D_{b,x,y}$, such that
\begin{equation}\label{dwdef}
    \sum_{b,x,y}D_{b,x,y}P(b|x,y)\leq w_d
\end{equation}
for all probabilities with a $d$-dimensional representation, while this
bound can be violated by a set of probabilities whose representation has a dimension strictly larger than $d$.
A family of dimension witnesses $I_N$ was
introduced recently ~\cite{gbha} in a scenario consisting of $N$
possible preparations and $N-1$ measurements with only two
possible outcomes, labeled by $b=\pm 1$. Here we shall focus on
one of these witnesses, namely
\begin{eqnarray}\label{Nm}
 I_4 &\equiv& E_{11}+E_{12}+E_{13} \nonumber \\
 &+& E_{21}+E_{22}-E_{23} \nonumber \\
 &+&E_{31}-E_{32} \nonumber \\
 &-&E_{41},
 \end{eqnarray}
where $E_{xy}=P(b=+1|x,y)-P(b=+1|x,y)$. Witness $I_4$ features
four preparations and three measurements. It can distinguish
ensembles of classical and quantum states of dimensions up to
$d=4$. All the relevant bounds are summarized in
Table~\ref{tableI4}.

\begin{table}
\begin{center}
  \begin{tabular}{||c|ccccc||}
  \hline\hline
  & \textrm{$C_2$ (bit)} & \textrm{$Q_2$ (qubit) }& \textrm{$C_3$ (trit)}& \textrm{$Q_3$ (qutrit)}& \textrm{$C_4$ (quart)}\\
    \hline
   $I_4$ & 5 & 6 & 7 & 7.97 & 9  \\
    \hline\hline
    \end{tabular}
\caption{\label{tableI4} Classical and quantum bounds for the
dimension witnesses $I_4$. The maximal possible value of $I_4$ for
classical systems of dimension $d$ is denoted $C_d$, while the
maximum possible value for quantum systems of dimension $d$ is
denoted $Q_d$. The witness can be used to discriminate ensembles
of classical and quantum states of dimension up to $d=4$. Note
that for a given value of the dimension of the system, the witness
can be used to certify its quantum nature. }
\end{center}
\end{table}


In order to test this witness experimentally, we must generate classical and quantum states of dimension 2 (bits and qubits, respectively), classical
and quantum states of dimension 3 (trits and qutrits), and classical states of dimension 4 (quarts). To do so we exploit the angular momentum of photons \cite{molina-terriza2007,mair2001}. The total angular momentum
contains a spin contribution associated with the polarization, and
an orbital contribution associated with the spatial shape of the
light intensity and its phase. Within the paraxial regime, both
contributions can be measured and manipulated independently. The
polarization of photons can be described in a $2$-dimensional
Hilbert space, as any polarization state can be described by a
weighted superposition of two orthogonal polarization states
(e.g., horizontal and vertical). The spatial degree of freedom of
light lives in an infinite-dimensional Hilbert space, as any
spatial waveform can be described in terms of a weighted
superposition of modes that span an infinite-dimensional basis
\cite{molina-terriza2002}. Paraxial Laguerre-Gaussian (LG) modes
constitute one of these bases. LG beams are characterized by two
integer indices, $p$ and $m$. The index $m$ determines the
azimuthal phase dependence of the mode, which is of the form $\sim
\exp \left( i m \varphi\right)$, where $\varphi$ is the azimuthal
angle in the cylindrical coordinates. LG laser beams carry a
well-defined orbital angular momentum (OAM) of $m\hbar$ per photon
that is associated with their spiral wavefronts \cite{allen1992}.

Our state preparator employs a source based on spontaneous
parametric down-conversion (SPDC) that generates pairs of photons
(signal and idler) entangled in both polarization and OAM. By
performing a projective measurement on one photon of the pair
(idler), we prepare its twin photon (signal) in a well-defined
state. A given measurement outcome on the first photon (idler)
corresponds to a given preparation for the second photon (signal),
which represents the mediating particle of Fig.~1.

\begin{figure}[t]
\centering
\includegraphics[width=1.0\columnwidth]{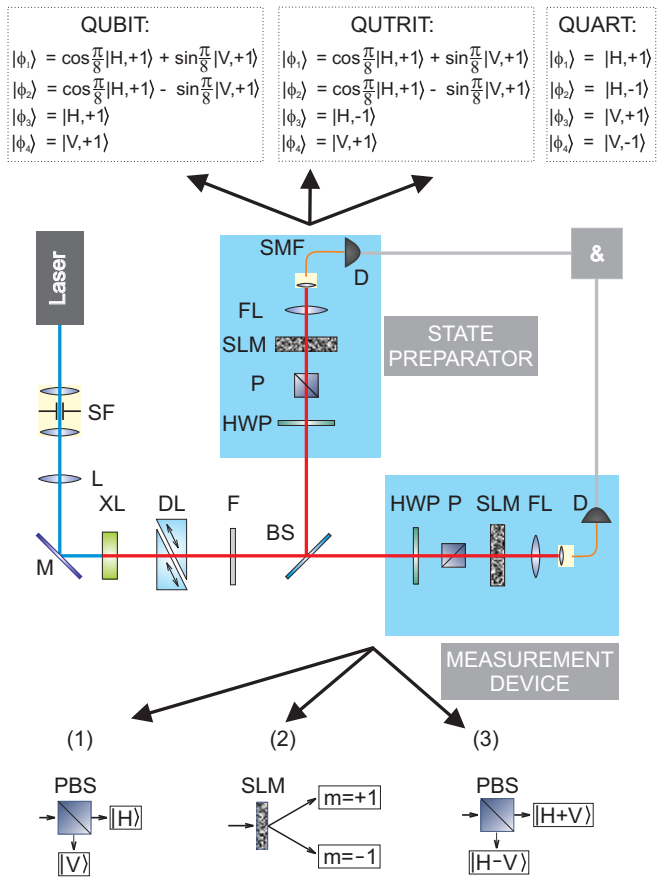}
 \caption{Experimental setup of the dimension witness. The upper part
depicts the individual states $\ket{\phi_x}$ ($x=1...4$) prepared
by the state preparator. The lower part illustrates the
measurements performed by the measurement device. SF - spatial
filter; L - lens; M - mirror; XL - nonlinear crystal; DL - delay
line; F - 5-nm bandpass filter; BS - beamsplitter; HWP - half-wave
plate; P - polarizer; SLM - spatial light modulator; PBS -
polarizing beamsplitter; FL - Fourier-transform lens; SMF -
single-mode fiber; D - single-photon counting detector; \& -
coincidence electronics.}
 \label{figure2}
\end{figure}

We use the state preparator to generate ensembles of states  of
dimension up to 4: two degrees of freedom are encoded in their
polarization (horizontal or vertical) and two degrees of freedom
are encoded in their OAM ($m = +1$ or $m = -1$). For the case of
qubit preparations, the four states $\ket{\phi_x}$ ($x=1...4$) are
enconded in the polarization only and have the same OAM of $m =
+1$ (see the upper left part of Fig.~\ref{figure2}). An analogous
procedure is used to prepare a qutrit state, however, now state
$\ket{\phi_3}$ carries OAM of $m = -1$, thus adding one dimension.
The quart state is prepared by flipping the OAM of state
$\ket{\phi_4}$. In this way we generate a 4-dimensional Hilbert,
spanned by the orthogonal vectors $\left\{\ket{H,+1}, \ket{H,-1},
\ket{V,+1},\rm{and} \ket{V,-1}\right\}$, where $\ket{H,\pm1}$ and
$\ket{V,\pm1}$ stand for a  horizontally or vertically polarized
photon with OAM equal to $m = \pm 1$, respectively.

To implement continuous transition from quantum to classical
states, a polarization-dependent temporal delay between the two
photons of an entangled pair is introduced. If the temporal delay
between the photons exceeds their correlation time, the coherence
is lost, i.e., the off-diagonal terms of the state in the ensemble
go to zero (see Appendix).

Finally, the measurement device simply performs measurements on
the second (signal) photon.

The experimental setup is shown in Fig.~\ref{figure2}. The second
harmonic (Inspire Blue, Spectra Physics/Radiantis) at a wavelength
of 405 nm of a Ti:saphire laser in the picosecond regime (Mira,
Coherent) is shaped by a spatial filter and focused into a 1.5-mm
thick crystal of beta-barium borate (BBO), where SPDC takes place.
The nonlinear crystal is cut for collinear type-II down-conversion
so that the generated photons have orthogonal polarizations.
Before splitting the photon pairs towards the state preparator and
measurement device, the polarization-dependent temporal delay
$\tau$ is introduced. The delay line (DL) consists of two quartz
prisms whose mutual position determines the difference between the
propagation times of photons with different polarizations.

\begin{figure}[t]
\centering
\includegraphics[height=0.8\columnwidth,angle=-90]{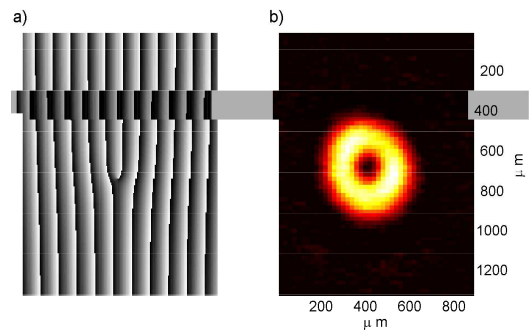}
 \caption{a) Computer generated hologram used to detect a LG beam with $m=+1$.
b) CCD camera photograph of the generated LG mode when the
hologram is illuminated by a Gaussian beam.} \label{holo}
\end{figure}

The measurement setup consists of a half-wave plate (HWP),
polarizer (P), spatial-light modulator (SLM) and a
Fourier-transform lens (FL). The half-wave plate and polarizer
project the incoming idler photon into the desired polarization
state. The desired OAM state is selected by the spatial light
modulator. If the nonlinear crystal is pumped by the fundamental
LG beam LG$_{00}$, about 15\% of the entangled photons are
generated in the LG mode with $m = \pm 1$ \cite{silvana}. The OAM
is conserved \cite{langford2004,molina-terriza2004,osorio2008}
(see Appendix for more details), i.e., if the signal photon has $m
= + 1$, the idler must have $m = - 1$ and vice versa. The SLM
encodes computer-generated holograms that transform the $m = +1$
state or $m = -1$ state into the fundamental LG state LG$_{00}$
\cite{molina-terriza2007} that is coupled into a single-mode fiber
(SMF). In this way, only a photon having a correct combination of
its OAM with a corresponding hologram placed at the SLM is
successfully coupled in the fiber and results in a subsequent
click at the detector. Figure~\ref{holo}~a) shows a
computer-generated hologram to measure an $m = + 1$ state and
Fig.~\ref{holo}~b) shows a picture of the LG mode taken by a CCD
camera when this hologram is illuminated by a Gaussian beam.

\begin{figure}[t]
\centering
\includegraphics[height=1.0\columnwidth,angle=-90]{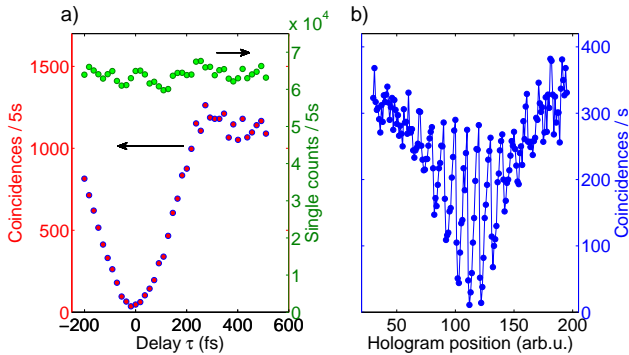}
 \caption{a) The Hong-Ou-Mandel dip shows the transition from a pure entangled state to a mixed state. The
pure state is used to prepare a quantum superposition of $\ket{H}$
and $\ket{V}$, whereas  the mixed state is used to prepare
statistical mixtures of $\ket{H}$ and $\ket{V}$. b)
Two-dimensional scan of the hologram position in the signal beam.
Signal hologram is set to $m = +1$ and idler hologram is set to $m
= 0$. Disappearance of coincidences asserts the correct position
of the hologram was found in order to detect signal photons with
OAM $m = + 1$.} \label{dip}
\end{figure}

When the polarizers in both arms are set to $+ 45$ degrees, a
Hong-Ou-Mandel dip \cite{HOM} of visibility above 95\% is measured
(see Fig.~\ref{dip}~a)). The center of the dip corresponds to the
pure state generation when the qubit and qutrit states are
prepared depending on their OAM. On the right-hand side, on the
shoulder of the dip, the superposition is lost and the qubit and
qutrit are converted to a classical bit and trit, respectively.
The purity of the state is given by Eq.~\ref{purity} that can be
found in Appendix with more details.

To find the correct position of the hologram on the SLM, the
center of the hologram must exactly coincide with the center of
the down-converted beam. To overlap them perfectly, first the
signal hologram is set to detect photons with $m = +1$ and the
idler hologram is set to detect $m = 0$. Then signal holograms
with different central positions of the vortex are scanned across
the SLM. Since the down-converted photons are generated with $m =
\pm 1$, a minimum of coincident counts between $m = +1$ and $m =
0$ tells us the correct position of the signal hologram (see
Fig.~\ref{dip}~b)). Now with the signal hologram fixed, the idler
hologram set to $m = -1$ is scanned to find a maximum number of
coincidences. With both holograms in the correct place, the
dimension witness is measured.

First, the qubit ensemble is generated. We sequentially prepare
the four qubit states $\ket{\phi_x}$ of Fig.~\ref{figure2} and
perform the three desired measurements. The first measurement $y =
1$ assigns dichotomic measurement results of $b = + 1$ and $b = -
1$ to horizontally and vertically polarized photons, respectively.
The second measurement $y = 2$ assigns $b = + 1$ and $b = - 1$ to
OAM values of $m = + 1$ and $m = - 1$, respectively. And the third
measurement $y = 3$ assigns $b = + 1$ and $b = - 1$ to photons
polarized at $+ 45^o$ and $- 45^o$. The maximum theoretical value
of the witness $I_4$ of Eq.~\ref{Nm} for this ensemble of states
and this set of measurements is 5.83 (see Appendix). The
experimentally measured witness reaches $I_4 = 5.66 \pm 0.15$.
This clearly demonstrates the quantum nature of our 2-dimensional
system, since classical bits always satisfy $I_4\leq5$.

\begin{figure}[t]
\centering
\includegraphics[height=0.9\columnwidth,angle=-90]{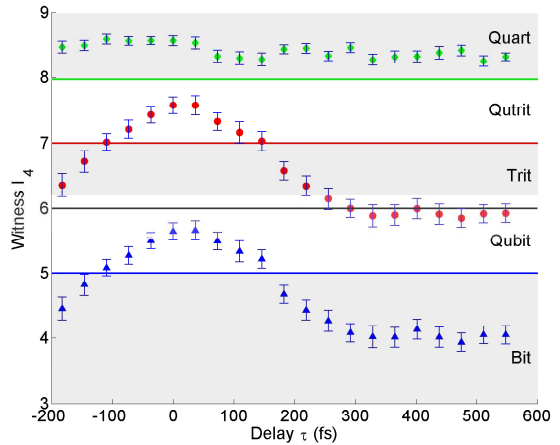}
 \caption{Dimension witness $I_4$ for qubit (blue triangles), qutrit (red circles)
and quart (green diamonds) as a function of temporal delay $\tau$.
For delays $\tau > 255$ fs, coherence is lost and quantum
superpositions turn into statistical mixtures, i.e., classical
states. The maximum violations for qubit, qutrit and quart are
$5.66 \pm 0.15$, $7.57 \pm 0.13$, and $8.57 \pm 0.06$,
respectively, which are close to the corresponding theoretical
bounds, given in Table~\ref{tableI4}, which are represented here
by the horizontal lines.} \label{figure3}
\end{figure}

Now a variable temporal delay $\tau$ is introduced between the
horizontal and vertical polarizations, and the value of the
witness decreases below 5, leveling off at $\tau \approx 255$ fs
that corresponds to the end of the Hong-Ou-Mandel dip, as expected
(see blue triangles of Fig.~\ref{figure3}). Thus our qubit is now
converted into a classical bit.

If the OAM of state $\ket{\phi_3}$ is flipped, a qutrit is
prepared and the dimension witness is measured. For a zero delay
$\tau$, the value of the witness reaches $I_4 = 7.57 \pm 0.13$,
certifying the presence of a (at least) 3-dimensional quantum
system. Here the maximum theoretical value for this ensemble of
states and this set of measurements is $5+2\sqrt{2} \sim 7.83$
(see Appendix).

When the delay line is scanned, the values of the witness drop
below 7, but in a certain range of $\tau$ they remain above the
qubit bound of 6, testifying that at least three dimensions are
present (see red circles of Fig.~\ref{figure3}). In the classical
limit of large delays, the witness values are still larger then
the bound of $I_4 = 5$ for bits.

Finally, we prepare classical 4-dimensional systems, i.e. quarts.
Now the first measurement $y = 1$ assigns the dichotomic value of
$b = + 1$ to horizontally polarized photons of $m = \pm 1$ and to
vertically polarized photons of $m = + 1$, and $b = - 1$
corresponds to vertically polarized photons with $m = -1$; the
second measurement $y = 2$ assigns $b = + 1$ and $b = - 1$ to
horizontally and vertically polarized photons, respectively; and
the third measurement $y = 3$ assigns $b = + 1$ and $b = - 1$ to
OAM values of $m = + 1$ and $m = - 1$, respectively. All the
correlation coefficients should add up and should be equal to 1.
The obtained value is $I_4 = 8.57 \pm 0.06$, which violates the
qutrit threshold by more than 10 standard deviations. In this
case, the values of the witness remain constant and are
independent of the temporal delay $\tau$. Indeed this is because
the state is classical (a statistical mixture of orthogonal
quantum states) and no superposition is present (see green
diamonds of Fig.~\ref{figure3}).

To conclude, we have demonstrated how the dimension of classical
and quantum systems can be bounded only from experimental
measurement results without any extra assumption on the devices
used in the experiment. Dimension witnesses represent an example
of a device-independent estimation technique, in which relevant
information about an unknown system is obtained solely from the
measurement data. Device-independent techniques (see also
Refs.~\cite{mayers,bardyn,bancal,rabelo}) provide an alternative
and robust approach to the estimation of quantum systems,
especially when compared to most of the existing estimation
techniques, such as quantum tomography, which crucially rely on
some assumptions about the measured system, e.g., its Hilbert
space dimension, that may be questionable in complex setups. Our
work demonstrates how the device-independent approach can be
employed to experimentally estimate the dimension of an unknown
system.

\section{Appendix}

In order to generate photons in the multidimensional Hilbert space
defined by the angular momentum of light, and at the same time to
be able to modify the dimension and the properties of the
corresponding quantum states, we generate photon pairs entangled
in the polarization and spatial degrees of freedom
\cite{mair2001}.

If the beam sizes of the pump beam and those of the collection
systems of the down-converted photons are large enough, we can
neglect the effects of the Poynting-vector walk-off that takes
place in a birefringent medium \cite{molina-terriza2007}. In this
case, the generated paired photons are found to be embedded into
LG modes with opposite winding numbers $m_s=-m_i$, where
$m_{s,i}\hbar$ refers to the OAM of the signal and idler photons,
respectively. After the down-conversion stage, a beam splitter is
used to separate the signal and idler photons. If the optical
system selects only the modes with $m_{s,i}=\pm 1$, the quantum
state of the photon pairs can be written as
\pagebreak
\begin{eqnarray}
\Psi &=&\frac{1}{\sqrt{2}}\int d\Omega \nonumber \\
& \times & \left\{ f\left(\Omega\right) \left[
\ket{\Omega,H,+1}_1 \ket{\Omega,V,-1}_2 \right. \right. \nonumber \\
& & \qquad \left. + \ket{\Omega,H,-1}_1 \ket{\Omega,V,+1}_2
\right] \nonumber \\
&+& f\left(-\Omega\right) \left[ \ket{\Omega,V,+1}_1
\ket{\Omega,H,-1}_2 \right. \nonumber \\
& & \qquad \left. \left. +\ket{\Omega,V,-1}_1 \ket{\Omega,H,+1}_2
\right] \right\}.
\end{eqnarray}
The signal photon has frequency $\omega_s=\omega_0+\Omega$, and
the idler photon has frequency $\omega_i=\omega_0-\Omega$, where
$\omega_0$ is the central frequency of the down-converted photons
and $\Omega$ is the frequency deviation from the central
frequency. The function $f$ reads
\begin{equation}
f(\Omega)=\frac{DL}{2\pi} \text{sinc} \left( \frac{D \Omega L}{2}
\right) \exp \left( i\frac{D \Omega L}{2} \right),
\end{equation}
where $L$ is the crystal length, and $D=1/v_s-1/v_i$ is the
difference of inverse group velocities ($v_{s,i}$) of the signal
and idler photons.

The signal and idler photons are entangled in the polarization and
spatial degrees of freedom. On the other hand, the frequency
properties of entangled two-photon states cannot be neglected even
when the entanglement resides in the other degrees of freedom. In
most experiments, the entangled states take advantage of only a
portion of the total two-photon quantum state and the generation
of high-quality entanglement requires the suppression of any
frequency ``which-path'' information that would otherwise degrade
the degree of entanglement. Notwithstanding, here we use the
presence of frequency ``which-path'' information to tailor the
quantum nature of ensembles of signal photons.

We introduce a temporal delay $\tau$ between the orthogonally
polarized signal and idler photons before they are separated at
the beam splitter. In the idler path, a polarizer rotated by an
angle $\varphi$, is placed, followed by a spatial light modulator
which in combination with a single mode fiber selects idler
photons in a LG mode with a winding number $m = \pm 1$. Thanks to
the angular momentum correlations between signal and idler
photons, the detection of an idler photon with horizontal or
vertical polarization ($\varphi=0^o$) and with a spatial shape
given by a LG mode with index $m=\pm 1$, projects the signal
photon into the orthogonal polarization and a LG mode with index
$m=\mp 1$. When the polarizer projects the idler photon in a
weighted combination of the two orthogonal polarizations ($\varphi
\ne 0^o$), the resulting quantum state of the signal photon
depends on the delay $\tau$.

The Hilbert space of signal photons, defined by the angular
momentum (i.e., polarization and orbital angular momentum), is
spanned by vectors $\left\{\ket{H,+1}, \ket{H,-1}, \ket{V,+1},
{\rm and} \ket{V,-1} \right\}$. After tracing out the frequency
degree of freedom, the quantum state of signal photons is given by
the density matrix
\begin{widetext}
\begin{equation}
\label{density_matrix} \rho_x=\left( \begin{array}{cccc}
  \alpha \cos^2 \varphi & 0  & \alpha \gamma(\tau) \sin 2\varphi/2 & 0 \\
 0 & (1-\alpha) \cos^2 \varphi & 0 & (1-\alpha) \gamma(\tau) \sin 2\varphi/2  \\
   \alpha \gamma^*(\tau) \sin 2\varphi/2 & 0 & \alpha \sin^2 \varphi  & 0 \\
  0 & (1-\alpha) \gamma^*(\tau) \sin 2\varphi/2  & 0 & (1-\alpha) \sin^2 \varphi,  \\
\end{array}
\right),
\end{equation}
\end{widetext}
where $\alpha=1$ if the idler photons is projected into a mode
with $m=-1$, and $\alpha=0$ if the idler photon is projected into
a mode with $m=1$.

The parameter $\gamma(\tau)$, which fulfills the condition
$|\gamma(\tau)|\le 1$, writes
\begin{eqnarray}
\gamma(\tau) &=& \frac{DL}{2\pi}\int d\Omega \text{sinc}^2
\frac{DL\Omega}{2} \exp \left\{ i\Omega \left(
2\tau-DL\right)\right\} \nonumber \\
&=& \text{tri}\left(\frac{\tau-DL/2}{DL/2} \right),
\end{eqnarray}
where tri(t) is the triangle function $\rm{tri}(t) =
max(1-|t|,0)$.

The purity of the state given by Eq.~(\ref{density_matrix}),
provided $\alpha=0,1$, writes
\begin{equation}
P=1-\frac{1-|\gamma(\tau)|^2}{2} \sin^2 2\varphi. \label{purity}
\end{equation}

The parameter $\gamma(\tau)$ defines the ``quantumness'' of the
ensemble of states. A $d$-dimensional ensemble of states $\left\{
\rho_x \right\}$ ($x=1 \ldots d$), where all the states satisfy
the condition $\gamma_x(\tau)=0$, constitutes an ensemble of
classical states, while if any of the states has $\gamma(\tau) \ne
0$, the ensemble is of quantum nature.

The quart is generated by the four combinations of $\alpha=0,1$
and $\varphi=0,90^o$. A three-dimensional quantum system (qutrit)
is generated by states $\ket{\phi_1}: \alpha = 1, \varphi =
22.5^o$, $\ket{\phi_2}: \alpha = 1, \varphi = -22.5^o$,
$\ket{\phi_3}: \alpha = 0, \varphi = 0^o$, and $\ket{\phi_4}:
\alpha = 1, \varphi = 90^o$. The delay line is set to $\tau=DL/2$
so that $|\gamma_x|=1$. When $|\gamma_x(\tau)| \rightarrow 0$, the
qutrit is converted into a classical trit. The theoretical value
of $I_4$ for this ensemble of states and this set of measurements
is
\begin{equation}
I_4=5+2\cos 2\varphi+2 \gamma(\tau) \sin2\varphi
\end{equation}
that attains its maximum value of $5+2\sqrt{1+|\gamma(\tau)|^2}$
when $\tan 2\varphi=\gamma(\tau)$. For $\gamma = 1$, we obtain
$I_4^{max} = 7.83 > 7$. For $\gamma = 0$, the maximum value a
classical trit can reach is 7.

A two-dimensional quantum system (qubit) is generated by the same
set of states as in the case of qutrit, except that state
$\ket{c}$ now has $\alpha=1$ and $\varphi=0^o$. A classical
two-dimensional system (bit) is obtained for $\gamma=0$. The value
of $I_4$ is
\begin{equation}
I_4=3+2\cos 2\varphi+2 \gamma(\tau) \sin2\varphi
\end{equation}
with its maximum of $3+2\sqrt{1+|\gamma(\tau)|^2}$. For $\gamma =
1$, we obtain $I_4^{max} = 5.83 > 5$ and for $\gamma = 0$ we
obtain $I_4^{max} = 5$.

\section{acknowldegments}
We acknowledge support from the ERC Starting Grant PERCENT, the EU
Projects Q-Essence and QCS, Spanish projects FIS2010-14830,
CatalunyaCaixa, and FI Grant of the Generalitat de Catalunya. This
work was also supported by projects FIS2010-14831, FET-Open grant
number: 255914 (PHORBITECH), and by Fundaci\'o Privada Cellex,
Barcelona.



\end{document}